\documentclass{aa}  

\usepackage{graphicx}
\usepackage{txfonts}
\usepackage{graphicx}	
\usepackage{amsmath}	
\usepackage{amsfonts}
\usepackage{multicol}   
\usepackage{bm}		    
\usepackage{pdflscape}	
\usepackage{acro}
\usepackage[normalem]{ulem}
\usepackage{natbib}
\usepackage{hyperref}
\usepackage[capitalise]{cleveref}
\usepackage{xcolor}
\usepackage{placeins}
\usepackage{afterpage}
\usepackage{longtable}

\newcommand{\iap}{CNRS \& Sorbonne Universit\'{e}, Institut d’Astrophysique de Paris (IAP), UMR 7095, 98 bis bd Arago, F-75014 Paris, France}
\newcommand{\cca}{Center for Computational Astrophysics, Flatiron Institute, 162 5th Avenue, New York, NY 10010, USA}
\newcommand{\oxford}{Astrophysics, University of Oxford, Denys Wilkinson Building, Keble Road, Oxford OX1 3RH, UK}
\newcommand{\icg}{Institute of Cosmology \& Gravitation, University of Portsmouth, Dennis Sciama Building, Portsmouth, PO1 3FX, UK}

\newcommand{\Mpch}{\ensuremath{h^{-1}\text{Mpc}}}
\newcommand{\camb}{\textsc{camb}}
\newcommand{\classcode}{\textsc{class}}
\newcommand{\bacco}{\textsc{bacco}}
\newcommand{\euclidemu}{\textsc{euclidemulator2}}
\newcommand{\halofit}{\textsc{halofit}}
\newcommand{\halofitplus}{\textsc{halofit+}}
\newcommand{\srhalofit}{\textsc{syren-halofit}}
\newcommand{\hmcode}{\textsc{hmcode}}
\newcommand{\colossus}{\textsc{colossus}}
\newcommand{\operon}{\textsc{operon}}
\newcommand{\quijote}{\textsc{quijote}}
\newcommand{\fortran}{\textsc{fortran90}}
\newcommand{\python}{\textsc{python3}}
\newcommand{\pico}{\textsc{pico}}
\DeclareMathOperator{\aq}{aq}
\newcommand{\hyper}[1]{\ensuremath{{}_{2}F_{1}}#1}

\newcommand{\splitatcommas}[1]{%
  \begingroup
  \begingroup\lccode`~=`, \lowercase{\endgroup
    \edef~{\mathchar\the\mathcode`, \penalty0 \noexpand\hspace{0pt plus 1em}}%
  }\mathcode`,="8000 #1%
  \endgroup
}

\begin{document} 

   \title{\srhalofit: A fast, interpretable, high-precision formula for the $\Lambda$CDM nonlinear matter power spectrum}

   \author{
    Deaglan J. Bartlett \thanks{\href{mailto:deaglan.bartlett@iap.fr}{deaglan.bartlett@iap.fr}} \inst{1}
    \and
    {Benjamin D. Wandelt} \inst{1,2}
    \and
    {Matteo Zennaro} \inst{3}
    \and
    {Pedro G. Ferreira} \inst{3}
    \and
    {Harry Desmond} \inst{4}
    }

    \institute{
        \iap
        \and
        \cca
        \and
        \oxford
        \and
        \icg
}

   \date{Received XXX; accepted YYY}

 
  \abstract
   {Rapid and accurate evaluation of the nonlinear matter power spectrum, $P(k)$, as a function of cosmological parameters and redshift is of fundamental importance in cosmology. Analytic approximations provide an interpretable solution, yet current approximations are neither fast nor accurate relative to numerical emulators.}
   {We aim to accelerate symbolic approximations to $P(k)$ by removing the requirement to perform integrals, instead using short symbolic expressions to compute all variables of interest. We also wish to make such expressions more accurate by re-optimising the parameters of these models 
   (using a larger number of cosmologies and focussing on cosmological parameters of more interest for present-day studies)
   and providing correction terms.}
   {
   We use symbolic regression to obtain simple analytic approximations to the nonlinear scale, $k_\sigma$, the effective spectral index, $n_{\rm eff}$, and the curvature, $C$, which are required for the \halofit{} model. We then re-optimise the coefficients of \halofit{} to fit a wide range of cosmologies and redshifts. We then again exploit symbolic regression to explore the space of analytic expressions to fit the residuals between $P(k)$ and the optimised predictions of \halofit. Our results are designed to match the predictions of \euclidemu{}, but we validate our methods against $N$-body simulations.
   }
   {
   We find symbolic expressions for $k_\sigma$, $n_{\rm eff}$ and $C$ which have root mean squared fractional errors of 0.8\%, 0.2\% and 0.3\%, respectively, for redshifts below 3 and a wide range of cosmologies. 
   We provide re-optimised \halofit{} parameters, which reduce the root mean squared fractional error (compared to \euclidemu{}) from 3\% to below 2\% for wavenumbers $k=9\times10^{-3}-9 \, h{\rm Mpc^{-1}}$.
   We introduce \srhalofit{} (symbolic-regression-enhanced \halofit), an extension to \halofit{} containing a short symbolic correction which 
   improves this error to 1\%.
   Our method is 2350 and 3170 times faster than current \halofit{} and \hmcode{} implementations, respectively, and 2680 and 64 times faster than \euclidemu{} (which requires running \classcode) and the \bacco{} emulator.
   We obtain comparable accuracy to \euclidemu{} and the \bacco{} emulator when tested on $N$-body simulations.
   }
   {
   Our work greatly increases the speed and accuracy of symbolic approximations to $P(k)$, making them significantly faster than their numerical counterparts without loss of accuracy.
   }

   \keywords{
   Cosmology: theory,
   Cosmology: cosmological parameters,
   Cosmology: large-scale structure of Universe,
   Methods: numerical
               }

    \titlerunning{\srhalofit: A symbolic nonlinear $P(k)$ emulator}
   \maketitle
%

\section{Introduction}

Under the current cosmological paradigm, the large-scale structure of the Universe evolved under gravity and cosmic expansion from highly Gaussian density fluctuations in the distant past to the present-day cosmic web. 
Despite the remarkable simplicity of the standard model ($\Lambda$CDM),  which contains just six parameters, the nonlinear equations of motion necessitate the computationally non-trivial task of simulating this evolution, which is typically done through expensive $N$-body simulations.
The nonlinear evolution has a dramatic effect on the matter power spectrum, $P(k)$, greatly enhancing the small-scale power compared to linear theory, and thus these effects cannot be neglected.
Even linear theory -- which is valid on large scales -- requires solving a hierarchy of coupled, linear differential equations \citep{Lewis_2000,Blas_2011,Hahn_2023}, demonstrating the difficulty in obtaining accurate predictions for $P(k)$.

Given the importance of the power spectrum in cosmological analyses, much effort has been put into by-passing these expensive simulations and directly predicting the matter power spectrum as a function of time and cosmological parameters.
Speed and differentiability of surrogate models are particularly attractive features. For example, the first high-precision emulator for the linear outputs of Boltzmann codes, \pico{} \citep{Pico2,Pico1} enabled the first use of Hamiltonian Monte Carlo methods for cosmological parameter inference \citep{Hajian_2007} due to these properties.

Approximations to the linear matter power spectrum are well-established, notably those of \citet{Eisenstein_1998,Eisenstein_1999}, which are accurate to a few percent, and the earlier work of \citet{Bardeen_1986}, which provided a less accurate approximation.
More recently, simple expressions which obtain a similar accuracy to \citeauthor{Eisenstein_1998} were obtained by \citet{Bayron-Orjuela-Quintana_2022,Bayron-Orjuela-Quintana_2023}, although these do not have the same physical motivation as the earlier works.
However, these symbolic expressions are insufficiently accurate for modern uses. This led \citet{Bartlett_2023_Pofk} to propose a simple extension to the \citeauthor{Eisenstein_1998} expressions which gives a root mean squared fractional error of just 0.2\% across a wide range of cosmologies, which is more than sufficient for current analyses \citep{Taylor_2018}.

On small scales, however, we must go beyond the linear predictions, and thus fitting formulae for the nonlinear matter power spectrum have proven essential. The scaling ansatz of \citet{Hamilton_1991}, which assumed that nonlinear structures decouple and form isolated systems, led to the development of the approximation of \citet{Peacock_1996}.
More recent approximations generically fall into one of three categories.

The first approach to predicting the nonlinear matter power spectrum is to use emulation techniques such as Neural Networks or Gaussian Processes, which are trained on $N$-body simulations to directly predict $P(k)$ given a set of cosmological parameters \citep{Heitmann_2009,Heitmann_2010,Heitmann_2014,Casarini_2016,Winther_2019,Angulo_2021,Arico_2021,Knabenhans_2021,SpurioMancini_2022,Mootoovaloo_2022,Zennaro_2023}.
The remaining two approaches are based on the halo model \citep{Ma_2000,Seljak_2000,Cooray_2002}, which assumes that the matter content of the Universe is bound in dark matter halos.
Under the approach adopted by \hmcode{} \citep{Mead_2015,Mead_2016,Mead_2021}, one can predict $P(k)$ by performing integrals of important quantities in the halo model, such as the density profile of halos and the halo mass function, which are assigned simple symbolic expressions.
In the \halofit{} model \citep{Smith_2003,Bird_2012,Takahashi_2012}, one assumes there are
two distinct contributions to the matter power spectrum; on large scales, $P(k)$ is dominated by the two-halo term, which describes correlations between the spatial positions of different halos, whereas on small scales, the one-halo regime, $P(k)$ is governed by the distribution of dark matter within individual halos. These two terms are approximated by analytic functions of the linear matter power spectrum and quantities derived from this (see \cref{sec:Halofit} for more detail).

The advantage of halo-model based approaches is that, involving only analytic expressions, they are much easier to interpret than numerical
emulators and have clearer extrapolation behaviour. Moreover, given the limited flexibility of their simple analytic expressions, 
the emulated predictions typically vary smoothly with input variables, so that the output is less noisy compared to a purely numerical approach \citep{Pico2}.
This is particularly desirable when utilising gradient-based optimisation or inference techniques since the derivatives of the analytic expressions will be smoother than their numerical counterparts.
Additionally, analytic expressions are highly portable since they can be easily implemented in the user's preferred programming language and do not become unusable when the codes underlying numerical approaches become deprecated.

However, in their current forms, there are two main disadvantages of the symbolic approaches compared to the emulators. Firstly, both symbolic approaches 
require one to perform integrals, which can be computationally slow.
This is a major problem for current analysis pipelines.
Second, they do not have the same level of accuracy as the numerical emulators.
The aim of this work is to solve both of these problems, in three steps:
\begin{enumerate}
    \item We produce simple analytic expressions (\cref{eq:ksigma_fit,eq:neff_fit,eq:C_fit}) for all variables appearing in the \halofit{} model so that, when coupled with the linear matter power spectrum approximation of \citet{Bartlett_2023_Pofk}, \halofit{} can be evaluated without performing any integrals, dramatically increasing its speed by a factor of 2350.
    \item We update the parameters of \halofit{} (\cref{eq:new_an,eq:new_bn,eq:new_cn,eq:new_gamman,eq:new_nu,eq:new_f1,eq:new_f2,eq:new_f3,eq:new_alphan,eq:new_betan})
    to optimise them for cosmologies relevant to present-day applications (see \cref{tab:cosmo_par_prior})
    This reduces the maximum error from $\sim10$\% for $k\lesssim 10 \, h {\rm Mpc^{-1}}$ to $\sim5$\%, when compared against the parameters of \citet{Takahashi_2012}
    across a wide range of cosmologies and redshifts.
    \item We provide a simple analytic correction (\cref{eq:A definition,eq:Halofit correction}) which can multiply the standard \halofit{} prediction to produce sub-percent level accurate power spectra. We note that (like all \halofit{} implementations but unlike the numerical emulators) a small residual remains due to imperfect modelling of the baryonic acoustic oscillations, but these are at approximately the percent level. This appears to be an inherent limitation of the \halofit{} formalism and should be addressed in future work.
\end{enumerate}

These significant improvements are enabled through the use of symbolic regression (SR) to rapidly and automatically search through the space of candidate fitting functions to yield high accuracy approximations of the various quantities of interest.
SR has gained popularity in recent years within the field of physics to 
re-discover laws of physics from data \citep{Lemos_2022},
propose candidate new ones \citep{Bartlett_2022,Bartlett_2023,Desmond_2023,Sousa_2023,Kamerkar_2022,Koksbang_2023c,Lodha_2023,Alestas_2022},
provide simple expressions to explain the output of complex simulations \citep{Delgado_2022,Miniati_2022,Koksbang_2023a,Koksbang_2023b,Wadekar_2022},
and obtain fitting functions for observed properties of astrophysical systems \citep{Russeil_2022,Russeil_2024}.

We note that in this paper we focus our attention on $\Lambda$CDM models only, whereas emulators such as \euclidemu{} and the \bacco{} emulator can handle non-zero neutrino mass and an evolving equation of state for dark energy. We leave such extensions to future work.

In \cref{sec:Theory} we provide theoretical background by presenting the \halofit{} model and describe the SR method we use to produce analytic approximations.
We obtain and present expressions for the variables required for \halofit{} in \cref{sec:SR halofit}, and we optimise the parameters of this model in \cref{sec:Optimise halofit}.
In \cref{sec:Correct halofit} we provide an extension to the traditional \halofit{} model -- which we call \srhalofit{} (symbolic-regression-enhanced \halofit) -- yielding percent level accurate predictions for $P(k)$.
Our expressions are validated against an independent set of simulations in \cref{sec:Performance}, where we also compare to existing emulators in both accuracy and speed. We conclude and
discuss future work in \cref{sec:Conclusions}.
Throughout this paper ``$\log$'' denotes the natural logarithm, and base-10 logarithms are denoted by ``$\log_{10}$''.

\section{Theoretical background}
\label{sec:Theory}

\subsection{Nonlinear matter power spectrum}
\label{sec:Halofit}

We wish to find an analytic approximation to the nonlinear matter power spectrum, $P(k;a,\bm{\theta})$, for a $\Lambda$CDM cosmology as a function of cosmological parameters, $\bm{\theta}$, scale factor, $a=1/(1+z)$, and wavenumber, $k$. 
In this paper we focus on the standard $\Lambda$CDM model, so $\bm{\theta}$ comprises of the (redshift zero) density parameter for baryons, $\Omega_{\rm b}$, and for all matter, $\Omega_{\rm m}$; the Hubble constant, $H_0 = 100 h {\rm \, km \, s^{-1} \, Mpc^{-1}}$; the scalar spectral index, $n_{\rm s}$; the curvature fluctuation amplitude, $A_{\rm s}$; and the reionisation optical depth, $\tau$.
Throughout this paper we use $\sigma_8$ instead of $A_{\rm s}$, which is defined as
\begin{equation}
    \label{eq:sigmaR}
    \sigma_R^2 = \int {\rm d}k \, \frac{k^2}{2 \pi^2} P (k; a=1, \bm{\theta}) \left| W(k,R) \right|^2,
\end{equation}
for
\begin{equation}
    W(k, R) = \frac{3}{(kR)^3} \left( \sin (k R) - kR \cos (k R) \right),
\end{equation}
where $\sigma_8$ is evaluated at $R = 8 \Mpch$.
Since $\tau$ has a very small effect on the power spectrum, we ignore this parameter throughout. We also set the neutrino mass to zero in all calculations.

In our previous work \citep{Bartlett_2023_Pofk}, we found an analytic emulator for the $z=0$ linear matter power spectrum. Given the zero-baryon Eisenstein \& Hu fit \citep{Eisenstein_1998,Eisenstein_1999}, $P_{\rm EH}(k; \bm{\theta})$, we defined 
\begin{equation}
    P_{\rm L}(k; a=1, \bm{\theta}) \equiv P_{\rm EH}(k; a=1, \bm{\theta}) F(k; \bm{\theta}),
\end{equation}
and found an expression for $F$ which was able to produce a sub-percent level approximation to $P_{\rm L}(k; a, \bm{\theta})$ for a range of cosmologies. Throughout this paper, when referring to this fitting formula, we mean the `fiducial' model given in \citet{Bartlett_2023_Pofk} as opposed to the more accurate yet less interpretable model given in the appendix of that paper.

Extending this to the nonlinear case requires addressing three differences compared to the linear theory prediction:
\begin{enumerate}
    \item $P(k;a,\bm{\theta})$ depends on the scale factor, $a$, in a non-trivial manner.
    \item Nonlinear physics smears the baryonic acoustic oscillation (BAO) peaks.
    \item Nonlinear physics boosts power on small scales (large $k$).
\end{enumerate}

To deal with these problems, we first note that it is trivial to scale the $z=0$ linear matter power spectrum to a different redshift:
\begin{align}
    P_{\rm EH}(k; a, \bm{\theta}) &= D(a, \bm{\theta})^2 P_{\rm EH}(k; a=1, \bm{\theta}),\\
    P_{\rm L}(k; a, \bm{\theta}) &= D(a, \bm{\theta})^2 P_{\rm L}(k; a=1, \bm{\theta}),
\end{align}
where the linear growth factor is (ignoring radiation)
\begin{equation}
    D(a, \bm{\theta}) \propto \hyper \left(\frac{1}{3}, 1, \frac{11}{6}, \frac{\Omega_{\rm m} - 1}{\Omega_{\rm m}} a^3 \right) a,
\end{equation}
where $\hyper$ is the Gauss hypergeometric function.

To address the remaining two problems, we base our approach on \halofit{} \citep{Smith_2003,Takahashi_2012}, a commonly used tool to model the nonlinear matter power spectrum. To do this, one defines a few variables based on the linear matter power spectrum. $k_\sigma$ is defined to be the wavenumber
at which
\begin{equation}
    \label{eq:ksigma definition}
    \sigma_{\rm G}^2(k_\sigma^{-1}) \equiv 1, \quad
    \sigma_{\rm G}^2(R) \equiv \int \Delta_{\rm L}^2 (k) \exp\left(-k^2 R^2\right) \, {\rm d} \log k,
\end{equation}
for
\begin{equation}
    \Delta_{\rm L}^2 (k) \equiv P_{\rm L}(k) \frac{k^3}{2 \pi^2},
\end{equation}
and $P_{\rm L}(k)$ is the linear matter power spectrum, where for brevity we suppress the dependence on $a$ and $\bm{\theta}$ for the remainder of this section.

Once we have $k_\sigma$, we also define
\begin{align}
    n_{\rm eff} + 3 &\equiv - \left. \frac{{\rm d} \log \sigma_{\rm G}^2(R)}{{\rm d} \log R} \right|_{\sigma_{\rm G} = 1},  \label{eq:neff definition} \\
    C &\equiv - \left. \frac{{\rm d}^2 \log \sigma_{\rm G}^2(R)}{{\rm d} \log R^2} \right|_{\sigma_{\rm G} = 1}. \label{eq:C definition}
\end{align}

Before continuing to describe the details of \halofit, it is worth emphasising how computationally demanding this is. First, one must run a Boltzmann solver to obtain the linear matter power spectrum. Then, one must run a root-finding algorithm to solve \cref{eq:ksigma definition} for $k_\sigma$, where each iteration involves performing an integral over $k$. Once one has identified $k_\sigma$, $\sigma_{\rm G}^2(R)$  must then be computed for many values of $R$ such that its first and second derivative at $k_\sigma$ can be calculated, either by finite differences or by fitting a spline. 
This is a large number of steps given that this is meant to be a fast method for approximating $P(k)$.

These variables are then combined to give the prediction for the matter power spectrum. Using the notation of \citet{Takahashi_2012}, we write
\begin{equation}
    \Delta^2(k) \equiv \Delta_{\rm Q}^2(k) + \Delta_{\rm H}^2(k), 
\end{equation}
where $\Delta_{\rm Q}^2(k)$ is the two-halo term and $\Delta_{\rm H}^2(k)$ is the one halo term.

\citet{Smith_2003} and \citet{Takahashi_2012} model the two-halo term as
\begin{equation}
    \label{eq:halofit delta q}
    \Delta_{\rm Q}^2(k) = \Delta_{\rm L}^2(k)
    \left[ \frac{\left( 1 + \Delta_{\rm L}^2(k) \right)^{\beta_n}}{1 + \alpha_n \Delta_{\rm L}^2 (k)} \right]
    e^{-f(y)},
\end{equation}
where $y \equiv k/k_\sigma$, $f(y) \equiv y/4 + y^2/8$, and the parameters $\alpha_n$ and $\beta_n$ are given by (fit from \citet{Takahashi_2012})
\begin{align}
    \alpha_n &= \left| 6.0835 + 1.3373 n_{\rm eff} - 0.1959 n_{\rm eff}^2 - 5.5274 C \right|, \\
    \beta_n &= 2.0379 - 0.7354 n_{\rm eff} + 0.3157 n_{\rm eff}^2 & \\
            & \quad + 1.2490 n_{\rm eff}^3 + 0.3980 n_{\rm eff}^4 - 0.1682 C.  \nonumber
\end{align}

They then model the one halo term as
\begin{align}
    \Delta_{\rm H}^2 (k) &= \frac{{\Delta_{\rm H}^\prime}^2 (k)}{1 + \mu_n y^{-1} + \nu_n y^{-2}}, \label{eq:halofit delta H}\\
    {\Delta_{\rm H}^\prime}^2 (k) &= \frac{a_n y^{3 f_1(\Omega_{\rm m})}}{1 + b_n y^{f_2(\Omega_{\rm m})} + \left[ c_n f_3(\Omega_{\rm m}) y\right]^{3 - \gamma_n}}.  \label{eq:halofit delta Hprime}
\end{align}
The parameters $a_n$, $b_n$, $c_n$, $\gamma_n$ and $\nu_n$ are, noting that $w=-1$ for our case, 

\begingroup
\allowdisplaybreaks
\begin{align}
\log_{10} a_n &= 1.5222 + 2.8553 n_{\rm eff} + 2.3706 n_{\rm eff}^2 & \\
                & \quad + 0.9903 n_{\rm eff}^3 + 0.2250 n_{\rm eff}^4 - 0.6038 C, \nonumber \\
    \log_{10} b_n &= -0.5642 + 0.5864 n_{\rm eff} + 0.5716 n_{\rm eff}^2 - 1.5474 C, \\
    \log_{10} c_n &= 0.3698 + 2.0404 n_{\rm eff} + 0.8161 n_{\rm eff}^2 + 0.5869 C, \\
    \log_{10} \gamma_n &= 0.1971 - 0.0843 n_{\rm eff} + 0.8460 C, \\
    \log_{10} \nu_n &= 5.2105 + 3.6902 n_{\rm eff}, \\
\end{align}
\endgroup
and the functions $f_i$ are given by
\begingroup
\allowdisplaybreaks
\begin{align}
    f_1 (\Omega_{\rm m}) &= \Omega_{\rm m}(z)^{-0.0307}, \\
    f_2 (\Omega_{\rm m}) &= \Omega_{\rm m}(z)^{-0.0585}, \\
    f_3 (\Omega_{\rm m}) &= \Omega_{\rm m}(z)^{0.0743},
\end{align}
\endgroup
where $\Omega_{\rm m}(z)$ is the total matter density parameter at redshift $z$.

\subsection{Symbolic regression}
\label{sec:SR}

To provide corrections to \halofit{}, we utilise the supervised machine learning technique of symbolic regression (SR) as implemented in the \operon{} package\footnote{\url{https://github.com/heal-research/operon}} \citep{Burlacu_2020}. \operon{} uses genetic programming \citep{turing,Goldberg,Affenzeller2009} to evolve a population of candidate mathematical expressions which attempt to fit a dataset given some inputs. Leveraging operations inspired by natural selection such as mutation and crossover (breeding), over several iterations the population evolves, with new members arising and poorly performing members being discarded. The expectation is that the population evolves to become fitter, and thus more accurate analytic expressions appear as the algorithm progresses.
\operon{} implements this procedure in a fast and memory efficient manner and has been shown to perform well in benchmark studies \citep{LaCava_2021,Burlacu_2023}.

All free parameters in expressions considered by \operon{} are optimised~\citep{Kommenda_2020} using the Levenberg–Marquardt algorithm \citep{Levenberg_1944,Marquardt_1963}. This includes additional scaling parameters, which appear at each terminal node in the expression tree, namely if $x$ is an input variable, then this will always appear as $c_i\times x$ for a parameter $c_i$, with a different value of $c_i$ for each occurrence of $x$. 
We define the model length to be the number of nodes appearing in an expression tree, excluding these scaling terms.
For example, the function $\cos(c_0 \times x)$ would have a model length of 2 and $\exp(c_0 \times x) + c_1 \times x$ has a model length of 4.

During non-dominated sorting, the objective values (root mean squared error and model length) of the equations are compared using the concept of $\epsilon$-dominance \citep{Laumanns_2002}.
There, a hyper-parameter $\epsilon$ is supplied to \operon{} such that two objective values within a distance $\epsilon$ of each other are considered equal, and is chosen to promote convergence of the algorithm. Different values of $\epsilon$ are used throughout this paper, which are chosen through experimentation with the aim of obtaining accurate yet compact expressions.

SR is a Pareto-optimisation problem due to the competing objectives of accuracy and simplicity: one can make an equation arbitrarily accurate by making it sufficiently complex, but making it too simple will give inaccurate predictions. However, one would expect that extremely complex models will have a tendency to over-fit the training data, so we use a validation set in all cases to aid this assessment.
When selecting models which balance the accuracy-simplicity trade-off, we cannot rely on principled statistical methods developed for SR \citep[see][]{Bartlett_2022,Bartlett_2023} due to the lack of measurement uncertainties in the quantities we wish to approximate. 
Instead, we begin by only considering those models lying on the Pareto front -- the most accurate model of a given length -- since these give the set of expressions which maximise one objective given a fixed value of the other.
We discard all functions which do not satisfy some predefined accuracy level and those for which the training and validation losses differ significantly. This leaves us with a set of models which we deem yield acceptable fits on both the training and validation datasets. We then choose the ``best'' model by visually inspecting the remaining expressions and making a subjective judgement based on their functional forms.

\section{Analytic approximations to \halofit{} variables}
\label{sec:SR halofit}

Rather than having to compute each of \cref{eq:ksigma definition,eq:neff definition,eq:C definition} on the fly (i.e. re-evaluate the linear matter power spectrum for the correct cosmology, perform a root-finding procedure, fit a spline and take derivatives), it would be preferable if each of these variables could be expressed as a simple function of cosmological parameters.

Unlike in \citet{Bartlett_2023_Pofk}, we cannot simply compute these quantities at redshift zero and re-scale, since all of $k_\sigma$, $n_{\rm eff}$ and $C$ are dependent on redshift in a non-trivial manner. 
To generate a training and validation set, we therefore sampled cosmologies from a Latin hypercube using the widths of parameters given in \cref{tab:cosmo_par_prior}, but added in an additional parameter, redshift. 
The range of the parameters are the same as those used to construct \euclidemu{} \citep{Knabenhans_2021} and those used in \citet{Bartlett_2023_Pofk}.
We sampled redshift uniformly in the range $[0,3]$, although we symbolically regressed in scale factor $a = 1 / (1 + z)$. We generated 200 training and 200 validation points which we fitted to.

We computed $k_\sigma$, $n_{\rm eff}$ and $C$ by performing the integral in \cref{eq:ksigma definition} using the linear power spectrum outputted by \camb{} \citep{Lewis_2000}. We could not use an infinite range of $k$ to perform this integral, but found that using a minimum $k$, maximum $k$ and logarithmic spacing of $k$ of
$10^{-4} \, h{\rm Mpc}^{-1}$, $10^2 \, h{\rm Mpc}^{-1}$ and $0.003$ (corresponding to 2000 $k$ values) gave converged results, where we integrated using Simpson's rule.

\begin{figure*}
  \centering
   \includegraphics[width=0.3\textwidth]{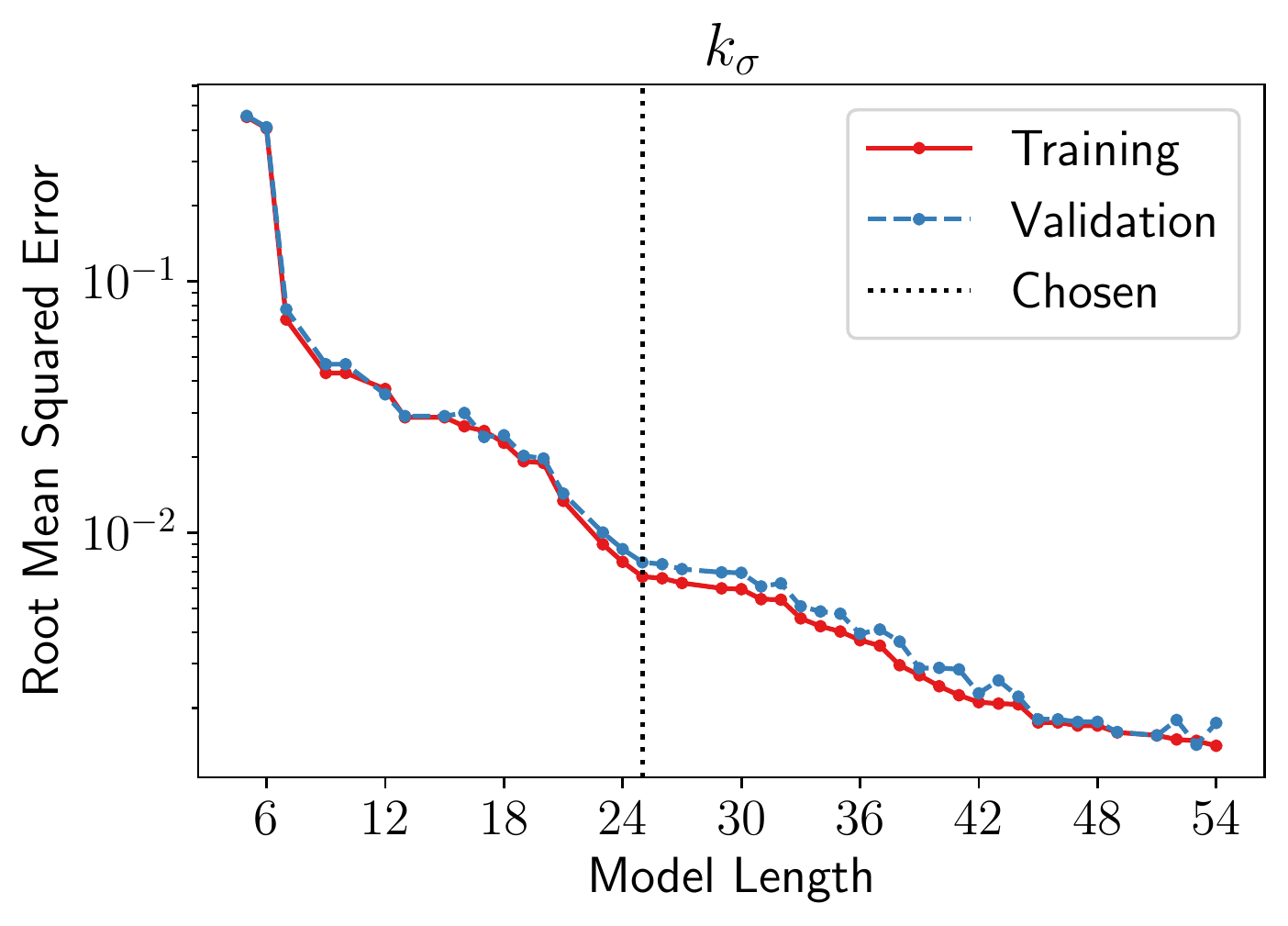}
   \includegraphics[width=0.3\textwidth]{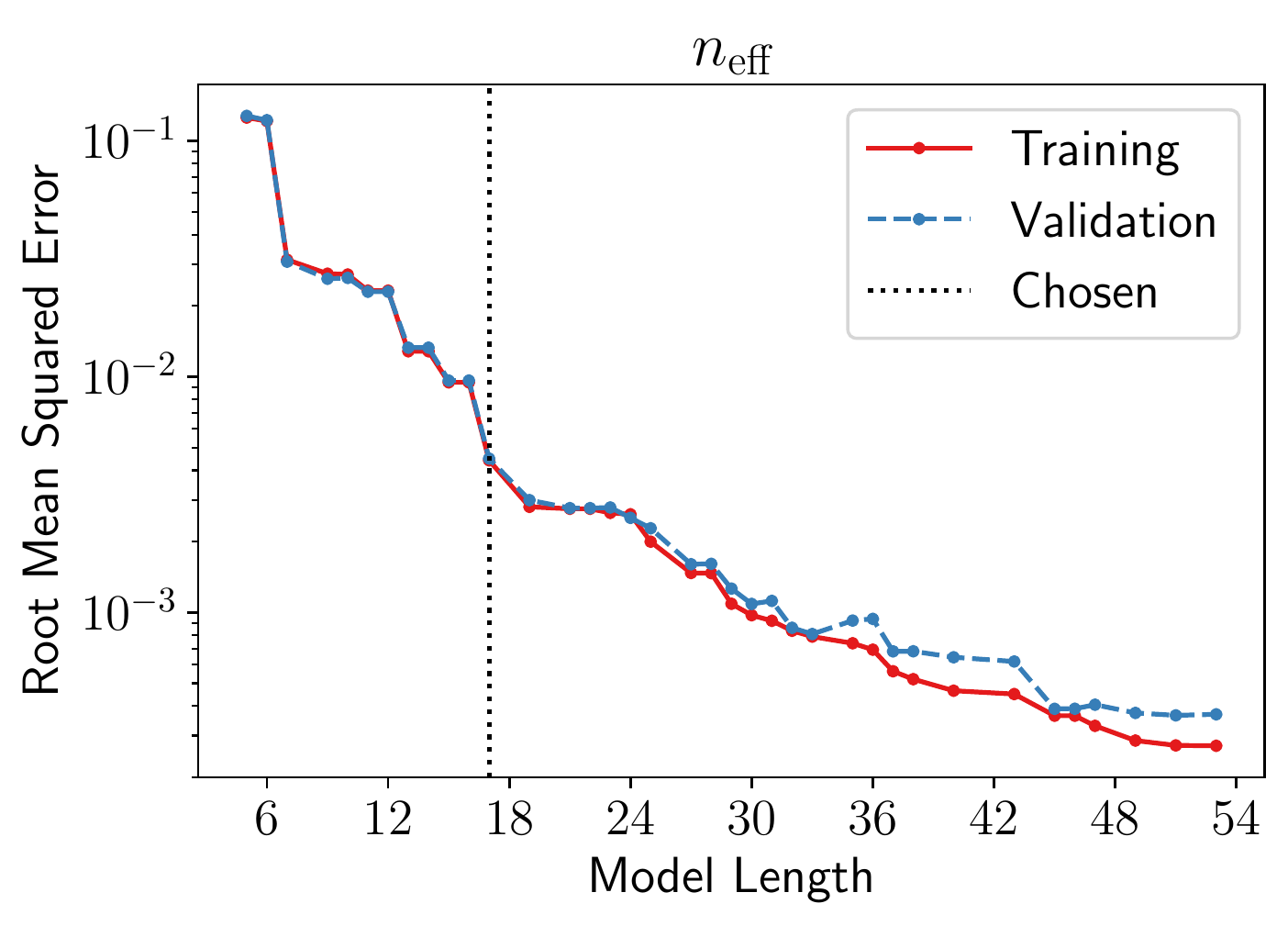}
   \includegraphics[width=0.3\textwidth]{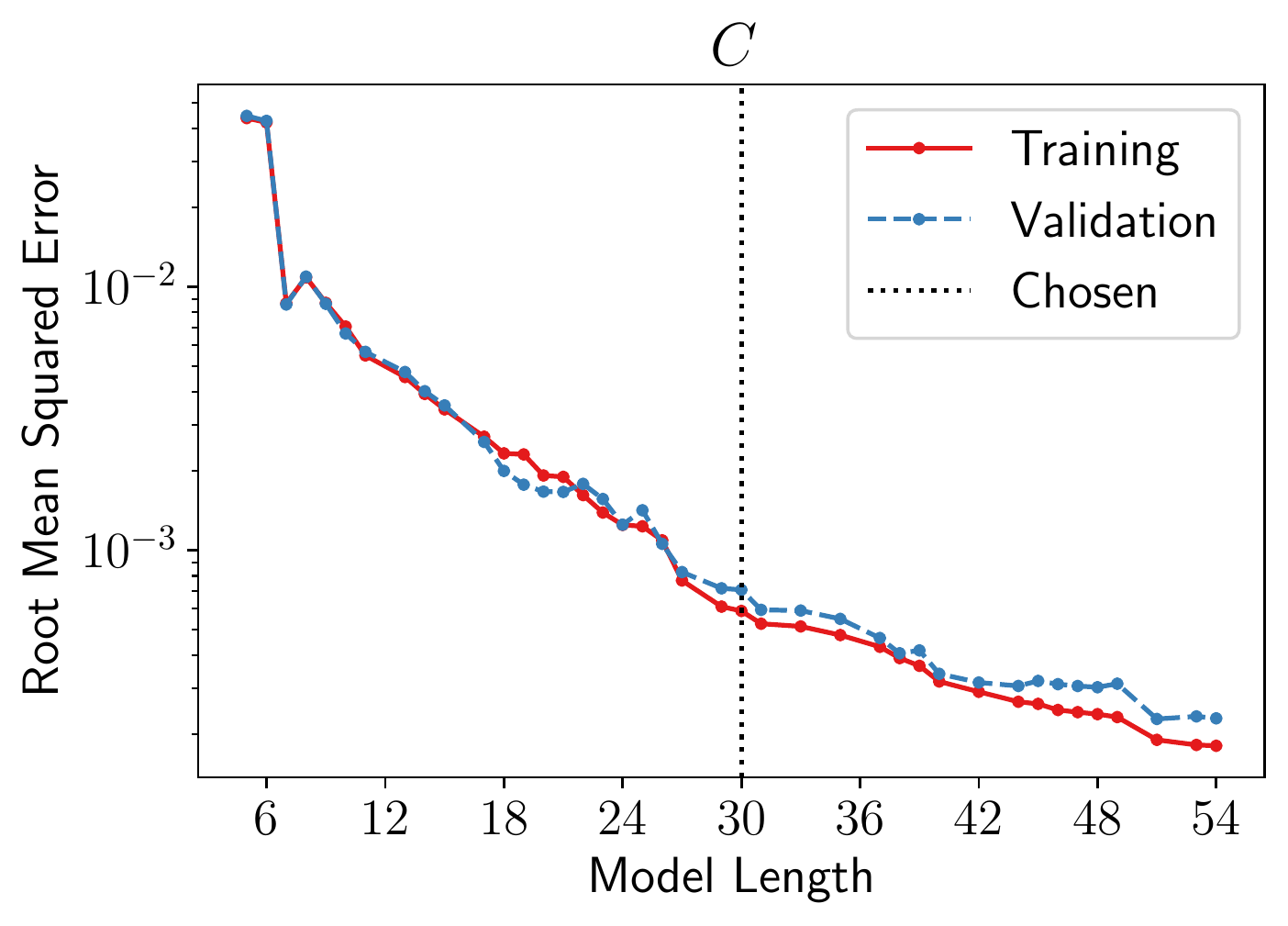}
  \caption{Pareto front of solutions found with \operon{} for $k_\sigma$ (left; \cref{eq:ksigma definition}), $n_{\rm eff}$ (centre; \cref{eq:neff definition}) and $C$ (right; \cref{eq:C definition}) over the range of cosmologies and redshifts considered. We plot the Pareto fronts for the training and validation data separately, indicating the lengths of our preferred models with vertical dotted lines.}
  \label{fig:halofit_pareto}  
\end{figure*}

We fitted each of $k_\sigma$, $n_{\rm eff}$ and $C$ as a function of cosmology using \operon, optimising simultaneously the root mean squared error (MSE) and the length of the model, using the basis operators
$+$, $-$, $\times$, $\div$, $\sqrt{\cdot}$, ${\rm pow}$ and $\log$.
The Pareto fronts of solutions found are given in \cref{fig:halofit_pareto}, with the settings described in the following sub-sections.

\begin{table}
    \centering
    \begin{tabular}{c|c|c}
        Parameter & Minimum & Maximum \\
        \hline\hline
         $10^9 \, A_{\rm s}$ & 1.7 & 2.5\\
         $\Omega_{\rm m}$ & 0.24 & 0.40 \\
         $\Omega_{\rm b}$ & 0.04 & 0.06 \\
         $h$ & 0.61 & 0.73 \\
         $n_{\rm s}$ & 0.92 & 1.00 \\
         $z$ & 0 & 3 \\
    \end{tabular}
    \caption{Range of cosmological parameters and redshift considered when constructing our analytic emulators. In all cases we generate samples in this six-dimensional space using a Latin hypercube, assuming a uniform distribution
    between the minimum and maximum values.}
    \label{tab:cosmo_par_prior}
\end{table}

\begin{figure*}
    \centering
    \includegraphics[width=\textwidth]{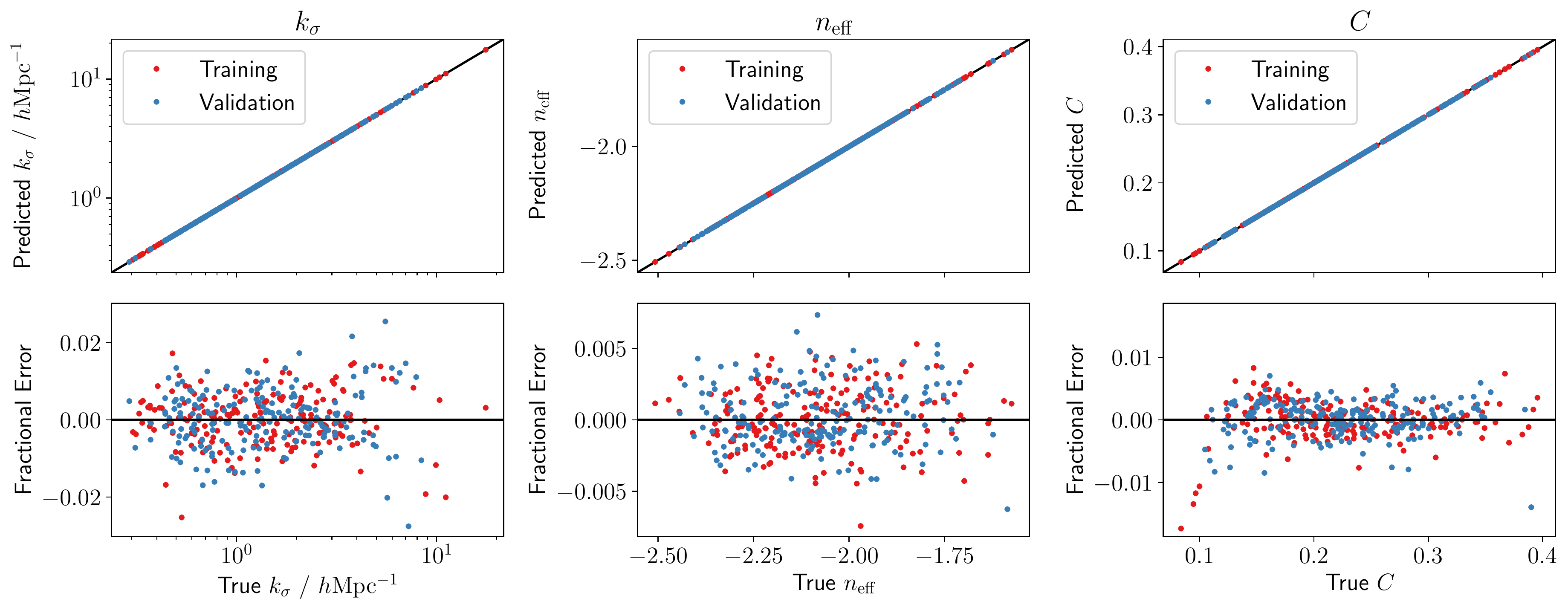}
    \caption{Predicted values (upper) and fractional errors (lower) for $k_\sigma$ (left; \cref{eq:ksigma definition}), $n_{\rm eff}$ (centre; \cref{eq:neff definition}) and $C$ (right; \cref{eq:C definition}) plotted against their true values, using the results in \cref{eq:ksigma_fit,eq:neff_fit,eq:C_fit}, respectively. The errors are almost always within 2\% for $k_\sigma$, 0.5\% for $n_{\rm eff}$ and 1\% for $C$.}
    \label{fig:halofit_parameters_residuals}
\end{figure*}

\subsection{Analytic approximation of \texorpdfstring{$k_\sigma$}{ksigma}}

Since at large redshifts the value of $k_\sigma$ can be large, we find that it is more appropriate to learn $\log k_\sigma$ rather than $k_\sigma$ directly. This has the advantage that $k_\sigma$ will always be predicted to be positive. We fitted this variable with \operon{} using $\epsilon = 10^{-5}$, running for 4 minutes or a maximum of $10^9$ function evaluations. 
From the resulting Pareto front, we see that there is a plateau in goodness of fit between
model lengths
$\sim$25-30. Inspecting these equations, we observed that the expression at model length 25 seems reasonable, so we chose this function. \operon{} gave an overparameterised form for this function and included an offset term which is much smaller than
the error on the fit, 
so can be neglected. Redefining some parameters and rearranging, we obtain 
\begin{equation}
    \label{eq:ksigma_fit}
    \begin{split}
         \log \left( \frac{k_\sigma}{h {\rm Mpc^{-1}}} \right) \approx &
        \frac{\psi_0}{\sigma_8 \left( a + \psi_9 n_{\rm s} \right)}
        \left[ \psi_1 a \left( \psi_2 - \sigma_8 \right) \right. \\
        & \left. + \left( \psi_3 a \right)^{-\psi_4 a - \psi_5 n_{\rm s}} \left( \psi_6 \Omega_{\rm b} + \left( \psi_7 \Omega_{\rm m} \right)^{-\psi_8 h} \right) \right],
    \end{split}
\end{equation}
where 
$\bm{\psi} = \{\splitatcommas{4.3576, 0.8358, 0.4302, 20.1077, 0.2593, 0.5732, 1.6809, 20.0433, 0.4257, 0.3908}\}$.

The fractional residuals on the training and validation sets are shown in the left panel of \cref{fig:halofit_parameters_residuals}, where we see an approximate percent-level agreement. The root mean squared fractional error on $k_\sigma$ is 0.7\% and 0.8\% for the training and validation sets, respectively.

\subsection{Analytic approximation of \texorpdfstring{$n_{\rm eff}$}{neff}}

Turning our attention to $n_{\rm eff}$, we again set the parameter $\epsilon=10^{-5}$ and ran for a maximum of 2 minutes or $10^8$ function evaluations. 
We see that there is a definitive kink in the Pareto front at a model length of around 17. Given that this is producing a root mean squared error of below $10^{-2}$ (in fact it is 0.2\% for both the training and validation sets) and it is not overly complicated when visually inspected, we chose this model. We note that \cref{fig:halofit_pareto} appears to show a significant drop in MSE at a model length of 19, but given the already low error and the appearance of a term containing $a^a$ in that expression, we chose the length-17 model. After some simplification, and noting again that the offset term produced by \operon{} is much smaller than the error so can be set to zero, this equation is
\begin{equation}
    \label{eq:neff_fit}
    n_{\rm eff} \approx 
    \left( \chi_0 n_{\rm s} - \chi_1 \right)
    \left( \chi_2 \Omega_{\rm b} - \chi_3 h + \left( \chi_4 a \right)^{- \chi_5 \Omega_{\rm m} - \chi_6 \sigma_8} \right),
\end{equation}
where 
$\bm{\chi} = \{ \splitatcommas{1.6514, 4.8815, 0.5125, 0.1488, 15.6499, 0.2393, 0.1346} \}$.
The residuals to this fit are given in the central panel of \cref{fig:halofit_parameters_residuals}, from which we see sub-percent fractional errors at all cosmologies and redshifts considered.

\subsection{Analytic approximation of \texorpdfstring{$C$}{C}}

Moving to a fit for $C$, we chose $\epsilon=10^{-3}$ (due to the smaller absolute values of $C$ compared to $\log k_\sigma$ and $n_{\rm eff}$) and ran for a slightly longer time than for $n_{\rm eff}$, with a maximum of 4 minutes or $10^9$ function evaluations. 
There is a slightly less noticeable kink in the Pareto front (\cref{fig:halofit_pareto}) than for $n_{\rm eff}$, although we see that after a model length of approximately 25, there is a reasonably sharp increase in accuracy, which then flattens up to a model length of approximately 35. Given the low errors in this regime, this seems a sensible place to choose a function from. Inspecting the models of these lengths, we found that many functions contain several nested exponential, which is undesirable. However, the model found with a length of 30 does not have this problem, so we chose this function. This is
\begin{equation}
    \label{eq:C_fit}
    \begin{split}
        C &\approx \varphi_{0} \sigma_{8} - \varphi_{1} \sqrt{\varphi_{2} n_{\rm s} + 
    \sigma_{8} \left(\varphi_{3} h - b_{6} + \left(\Omega_{\rm m} \varphi_{4}\right)^{a \varphi_{5}}\right)} \\
    & \quad \times \left(\Omega_{\rm b} \varphi_{7} + a \varphi_{8} + \varphi_{9} \sigma_{8} - \left(\varphi_{10} h\right)^{\Omega_{\rm m} b\varphi_{11}}\right) - \varphi_{12},
    \end{split}
\end{equation}
where 
$\bm{\varphi} = \{ \splitatcommas{0.3359, 1.4295, 0.1153, 0.0572, 48.0722, 0.1941, 1.176, 1.0151, 0.2354, 0.3596, 2.3898, 0.3569, 0.4431} \}$.

The fractional residuals of this fit are shown in the right panel of \cref{fig:halofit_parameters_residuals}, where we see that we are always accurate within 2\%, but almost always within 0.5\%. This is reflected in the root mean squared fractional error on $C$ of 0.3\% for both the training and validation sets.

\section{Optimised \halofit{} parameters}
\label{sec:Optimise halofit}

Now we have symbolic expressions for $k_\sigma$, $n_{\rm eff}$ and $C$, we can rewrite \halofit{} using (1) the \citet{Bartlett_2023_Pofk} linear $P(k)$ emulator and (2) the newly constructed emulators for these variables. Before adding any additional terms, we re-optimised the parameters of \halofit{} using the emulated values of $P_{\rm L}$, $k_\sigma$, $n_{\rm eff}$ and $C$. 
The commonly used parameters of \citet{Takahashi_2012} were optimised considering sixteen sets of cosmological parameters: six cosmologies based on the results of \textit{WMAP} \citep{Spergel_2003,Spergel_2007,Komatsu_2009,Komatsu_2011}, and ten taken from the Coyote universe project \citep{Heitmann_2009,Heitmann_2010,Lawrence_2010}. Since then, improved constraints on cosmological parameters mean these may not cover the region of parameter space which is of interest for present-day experiments, with some falling outside the range given by \cref{tab:cosmo_par_prior}. 
Thus, we re-optimised the \halofit{} parameters so that they provide the best possible fits across the range of parameters given in \cref{tab:cosmo_par_prior}.

To perform this optimisation, we need access to the nonlinear matter power spectrum. To allow us to use any redshift, we generated $P(k;a,\bm{\theta})$ using \euclidemu{} \citep{Knabenhans_2021}, a fast and accurate method for predicting the ratio of the nonlinear matter power spectrum to the linear one. We chose to do this rather than generate new $N$-body realisations since such a generation is much more computationally expensive. \euclidemu{} is accurate to the percent level, so we can achieve a high degree of accuracy from fitting the emulated spectra, and this allows more flexibility in choosing redshift values. We validate our results against $N$-body simulations in \cref{sec:Accuracy test}.

We again used 200 cosmologies drawn from a Latin hypercube with the same bounds as in \cref{sec:SR halofit} (including the addition of redshift). We used 100 logarithmically spaced $k$ values in the range $9 \times 10^{-3} - 9 \, h{\rm Mpc^{-1}}$.
Specifically, we optimised the parameters of $a_n$, $b_n$, $c_n$, $\gamma_n$, $\nu_n$, $\alpha_n$ and $\beta_n$ and the exponents of $f_1$, $f_2$ and $f_3$. As a loss function, we used the mean fractional squared error on the \halofit{} model compared to \euclidemu{}:
\begin{equation}
    \label{eq:halofit_loss}
    L = \sum_{i, j} \left( \frac{P_{\rm halofit}\left(k_i; a_j, \bm{\theta}_j \right) -  P_{\rm euclid} \left(k_i; a_j, \bm{\theta}_j \right)}{P_{\rm euclid}\left(k_i; a_j, \bm{\theta}_j \right)} \right)^2,
\end{equation}
where $i$ indexes the 100 values of $k$ and each point on the Latin hypercube is given by scale factor $a_j$ and cosmological parameters $\bm{\theta}_j$.

We optimised \cref{eq:halofit_loss} using an Adam optimiser \citep{Kingma_2014}. The initial learning rate was set to 0.01, and we used the StepLR learning rate scheduler, with a step size of 100 and gamma=0.9. We ran the optimiser for $2\times10^4$ iterations, but found that it converged much before this. As an initial guess, we use the parameters used in \citet{Takahashi_2012}.

\begin{figure*}
    \centering
    \includegraphics[width=\textwidth]{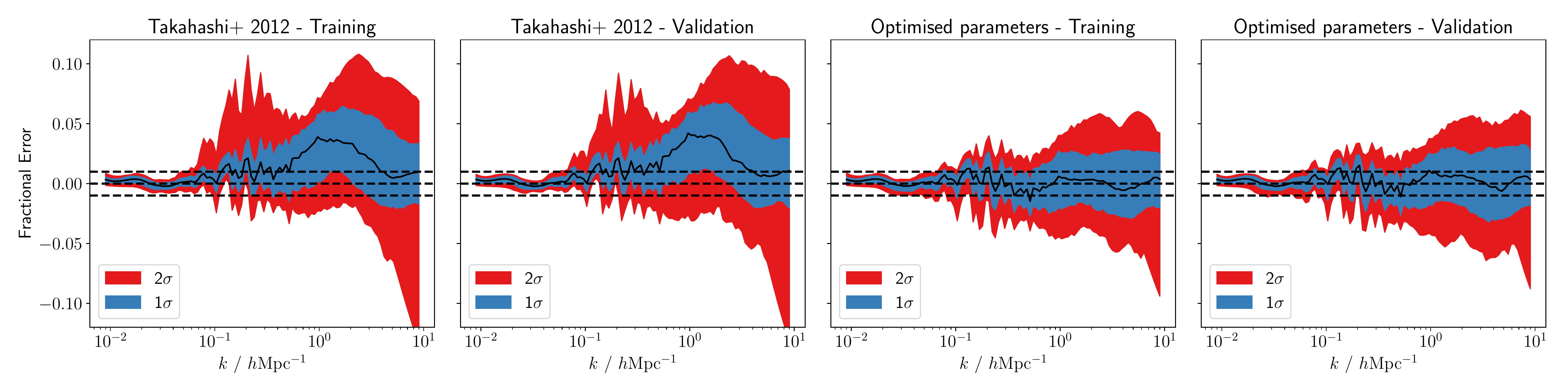}
    \caption{Comparison between the \citet{Takahashi_2012} \halofit{} parameters and the new optimised results (\halofitplus), with the bands giving the 1 and 2$\sigma$ errors, where we assume that the result of \euclidemu{} is the truth. For both training and validation we use 200 cosmologies and 100 values of $k$. 
    The dashed horizontal lines indicate an error of $\pm$1\%.
    The new parameters dramatically reduce the errors, particularly 
    for $k \gtrsim 10^{-1} \, h {\rm Mpc^{-1}}$.}
    \label{fig:new_halofit_errors}
\end{figure*}

After optimising the parameters and rounding to four decimal places, we obtained new expressions for the \halofit{} parameters:
\begingroup
\allowdisplaybreaks
\begin{align}
	\log_{10} a_n &= 1.5358 + 2.8533 n_{\rm eff} + 2.3692 n_{\rm eff}^2 \label{eq:new_an}  \\
                        & \quad + 0.9916 n_{\rm eff}^3 + 0.2244 n_{\rm eff}^4 - 0.5862 C, \nonumber \\
	\log_{10} b_n &= -0.5650 + 0.5871 n_{\rm eff} + 0.5757 n_{\rm eff}^2 - 1.5050 C, \label{eq:new_bn} \\
	\log_{10} c_n &= 0.3913 + 2.0252 n_{\rm eff} + 0.7971 n_{\rm eff}^2 + 0.5989 C, \label{eq:new_cn} \\
	\gamma_n &= 0.2216 - 0.0010 n_{\rm eff} + 1.1771 C, \label{eq:new_gamman}  \\
	\log_{10} \nu_n &=  5.2082 + 3.7324 n_{\rm eff}, \label{eq:new_nu} \\
	f_1 \left( \Omega_{\rm m} \right) &= \Omega_{\rm m}(z)^{-0.0158}, \label{eq:new_f1} \\
	f_2 \left( \Omega_{\rm m} \right) &= \Omega_{\rm m}(z)^{-0.0972}, \label{eq:new_f2} \\
	f_3 \left( \Omega_{\rm m} \right) &= \Omega_{\rm m}(z)^{0.1550},  \label{eq:new_f3} \\
	\alpha_n &= \left| 6.1043 + 1.3408 n_{\rm eff} - 0.2138 n_{\rm eff}^2 - 5.3250 C \right|, \label{eq:new_alphan}  \\
	\beta_n &= 1.9967 - 0.7176 n_{\rm eff} + 0.3108 n_{\rm eff}^2 \label{eq:new_betan} \\
                & \quad + 1.2477 n_{\rm eff}^3 + 0.4018 n_{\rm eff}^4 - 0.3837 C. \nonumber
\end{align}
\endgroup
For the remainder of the paper, we refer to the \halofit{} model with these parameters as \halofitplus.
We note that these parameters are not dramatically different to \citet{Takahashi_2012}, so this represents a small refinement of their fit.
We compare the results of this choice of parameters to those of \citet{Takahashi_2012} in \cref{fig:new_halofit_errors}, where we see a particularly dramatic improvement in the one-to-two halo regime. The root mean squared fractional error for \citet{Takahashi_2012} is 2.9\% and 3.0\% on the training and validation sets, respectively, whereas this is reduced to 1.8\% and 1.9\%, respectively, with the optimised parameters.
The maximum error for $k\lesssim 10 \, h {\rm Mpc^{-1}}$ is also reduced, from $\sim10$\% to $\sim5$\%.

\section{Corrections to \halofit}
\label{sec:Correct halofit}

Up to this point, we have not modified the functional form of \halofit. Instead, we have made two major improvements. First, we have increased its speed by removing the requirement to run a root-finding algorithm on integrals of the power spectrum, and having to compute derivatives of the result, by producing concise and accurate symbolic expressions for the output of this procedure. Second, by re-optimising the coefficients appearing in the \halofit{} formalism, we have improved the accuracy of the method, with root mean squared errors of better than 2\%. 
In this section, we provide a correction to
the functional form of
\halofit{} that will allow us to obtain sub-percent level errors. 

We parameterise this additional term as
\begin{equation}
    \label{eq:A definition}
    P (k; a, \bm{\theta}) = \left(  P_{\rm Q} (k; a, \bm{\theta}) + P_{\rm H} (k; a, \bm{\theta}) \right) \left( 1 + A(k; a, \bm{\theta}) \right),
\end{equation}
where we expect that $A \ll 1$ (we need to provide at most a correction at the level of a few percent) and that $A \to 0$ as $k\to 0$ so that the linear power spectrum is realised on large scales.

To find an expression for $A$, we again generated training and validation data on Latin hypercubes, using different seeds to those used in the previous sections but the same parameters. We again chose to sample 200 cosmologies for both training and validation and use 100 logarithmically spaced $k$ values between $9\times 10^{-3}$ and $9 \, h{\rm Mpc^{-1}}$. We computed the 
\halofitplus{}
prediction using the analytic expressions for $k_\sigma$, $n_{\rm eff}$ and $C$ obtained in \cref{sec:SR halofit}, the analytic approximation to the linear matter power spectrum of \citet{Bartlett_2023_Pofk}, and the parameters found in \cref{sec:Optimise halofit}. We divided the output of \euclidemu{} by this quantity and subtracted 1 to obtain the target values of $A$. To make our target $\mathcal{O}(1)$, we chose to fit for $10A$ rather than $A$.

Again, we used \operon{} to obtain candidate analytic expressions. We found that a value of $\epsilon=10^{-3}$ is appropriate and ran for 24 hours using a single node with 128 cores, optimising root mean squared error and model length. Since we do not want very complex expressions, we limited the maximum model length to 70. All candidate expressions are comprised of standard arithmetic operations (addition, subtraction, multiplication, division), as well as the natural logarithm, square root, cosine, power and analytic quotient operator ($\aq(x,y) \equiv x / \sqrt{1+y^2}$). 
We found that using $y=k/k_\sigma$ enabled a more efficient search than using $k$ directly, so we attempted to find $A$ as a function of $y$, $k_\sigma$, $n_{\rm eff}$, $C$, $a$ and the cosmological parameters.

\begin{figure}
    \centering
    \includegraphics[width=\columnwidth]{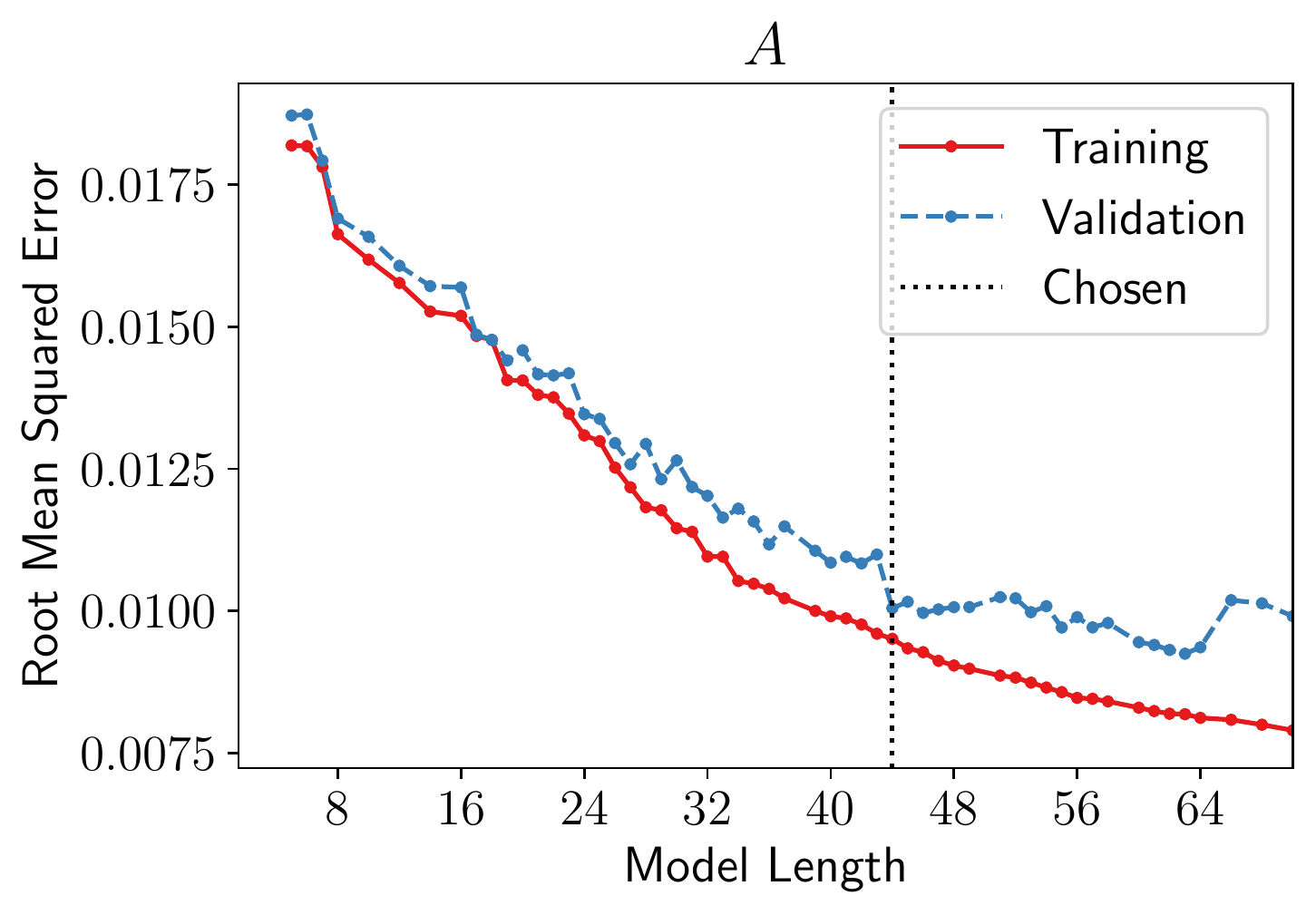}
    \caption{Pareto front of solutions found by \operon{} to approximate the difference between the nonlinear matter power spectrum and the prediction of \halofit{}. Each point on the red line represents the function with the best mean squared error on the training set for a given model length, whereas the blue curve shows the same loss for these functions evaluated on the validation set. We choose to use the model of length 44, as indicated by the vertical dotted line.}
    \label{fig:A_pareto}  
\end{figure}

In \cref{fig:A_pareto} we plot the Pareto front of solutions obtained by \operon{} which approximate $A$. We see that for model lengths up to approximately 25, the training and validation losses are approximately equal. Beyond this point, although both losses continue to improve, the rate of improvement decreases for the validation loss, until it stalls at a model length of 44, suggesting a degree of overfitting beyond this point. We therefore chose this function as our approximation to $A$, which, after merging superfluous constants in the expression, is
\begin{equation}
    \label{eq:Halofit correction}
    \begin{split}
           & A \approx
           - \frac{d_0}{\sqrt{1 + \left( d_1 y \right)^{-d_2 C}}}
           \Biggl[
           y
           - \frac{d_3 \left(y - d_4 n_{\rm s} \right)}{\sqrt{\left(y - d_5 \log \left( d_6 C\right)\right)^2 + d_7}} \\
           & +
           \frac{d_8 n_{\rm eff}}{\sqrt{ \left( d_{9}^2 + \sigma_8^2 \right) \left( \left( d_{10} y - \cos \left( d_{11} n_{\rm eff} \right) \right)^2 + 1 \right)}} \\
           & + 
           \frac{\left( d_{12} + d_{13} n_{\rm eff} - d_{14} C - d_{15} y \right) \left( d_{16} n_{\rm eff} + d_{17} y + \cos \left( d_{18} n_{\rm eff} \right)\right)}{\sqrt{y^2 + d_{19}}}
           \Biggl],
    \end{split}
\end{equation}
where the parameters, $\{d_i\}$, are tabulated in \cref{tab:A_params}. 
For the remainder of the paper, we refer to the combination of this correction to \halofit{} model with parameters given by \cref{sec:Optimise halofit} as \srhalofit, the primary product of this work.

Upon inspecting \cref{eq:Halofit correction}, one notices that, although we allowed the expression to depend on all of the cosmological parameters, the only parameters which appear beyond the standard halofit variables are $n_{\rm s}$ and $\sigma_8$. It is perhaps unsurprising that these are the two most important variables given that they characterise the slope and normalisation of the initial power spectrum, although it is interesting to find that no other variables are required.

We note that \operon{} provided an additional constant offset for \cref{eq:Halofit correction}, but this is much smaller than the values of $A$ produced, so we ignore it. In fact, this is justified because the leading order term of \cref{eq:Halofit correction} as $y \to 0$ is $\mathcal{O}(y^{d_2 C / 2})$ which, given that $d_2 > 0$ and $C > 0$ for all cosmologies considered, tends to 0 as $y \to 0$. Hence \cref{eq:Halofit correction} does not produce any correction to \halofit{} on large scales, which is desirable as we know linear theory should hold for small $k$, and this is already the limiting behaviour of 
\halofitplus.
Keeping the constant offset would not allow this to hold exactly.
This demonstrates the advantage of a symbolic expression for $P(k)$ over numerical emulators: the extrapolation behaviour beyond the range of the training data is clear
and can be controlled, 
so we can ensure that the model behaves as desired.

\begin{table}[]
	\centering
	\begin{tabular}{c|l|c|l}
		Parameter & Value &Parameter & Value \\
		\hline\hline
		 $d_{0}$ &  $0.2011$ & $d_{10}$ &  $0.3377$ \\
		 $d_{1}$ &  $1.2983$ & $d_{11}$ &  $3.3150$ \\
		 $d_{2}$ &  $16.8733$ & $d_{12}$ &  $3.9819$ \\
		 $d_{3}$ &  $3.6428$ & $d_{13}$ &  $1.3572$ \\
		 $d_{4}$ &  $1.0622$ & $d_{14}$ &  $3.3259$ \\
		 $d_{5}$ &  $0.1023$ & $d_{15}$ &  $0.3872$ \\
		 $d_{6}$ &  $2.2204$ & $d_{16}$ &  $4.1175$ \\
		 $d_{7}$ &  $0.0105$ & $d_{17}$ &  $2.6795$ \\
		 $d_{8}$ &  $0.4870$ & $d_{18}$ &  $5.3394$ \\
		 $d_{9}$ &  $0.6151$ & $d_{19}$ &  $0.0338$ \\
	\end{tabular}
	\caption{Best-fit parameter values for the correction to \halofit{} given in \cref{eq:Halofit correction}. All parameters are dimensionless.}
	\label{tab:A_params}
\end{table}

\begin{figure*}
    \centering
    \includegraphics[width=\textwidth]{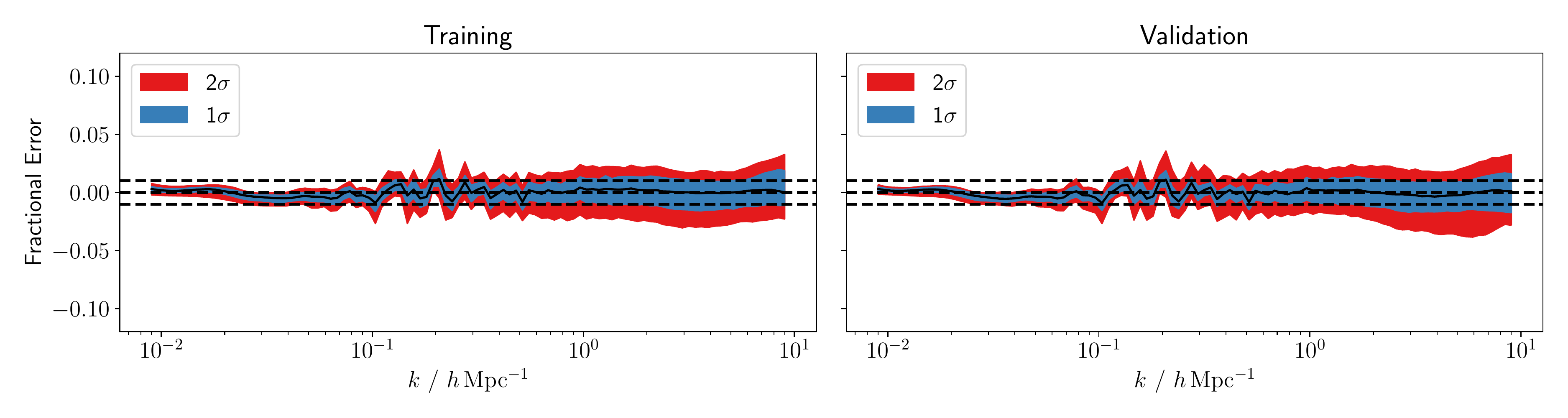}
    \caption{Distribution of fractional differences between \euclidemu{} and the prediction from \halofit{} plus the correction given in \cref{eq:Halofit correction} (\srhalofit). For ease of comparison, the range of the $y$ axis is the same as \cref{fig:new_halofit_errors}. The bands give the 1 and $2\sigma$ values, and we find that the root mean squared fractional error is 0.9\% and 1.0\% for training and validation, respectively.}
    \label{fig:residuals_pk_nonlinear}
\end{figure*}

We plot the fractional differences between the nonlinear power spectrum produced by \euclidemu{} and that from \srhalofit{}
in \cref{fig:residuals_pk_nonlinear}.
The errors are now reduced to a root mean squared fractional error of 0.9\% and 1.0\% for the training and validation sets, respectively, and a mean absolute fractional error of 0.7\%.

\begin{figure}
    \centering
    \includegraphics[width=\columnwidth]{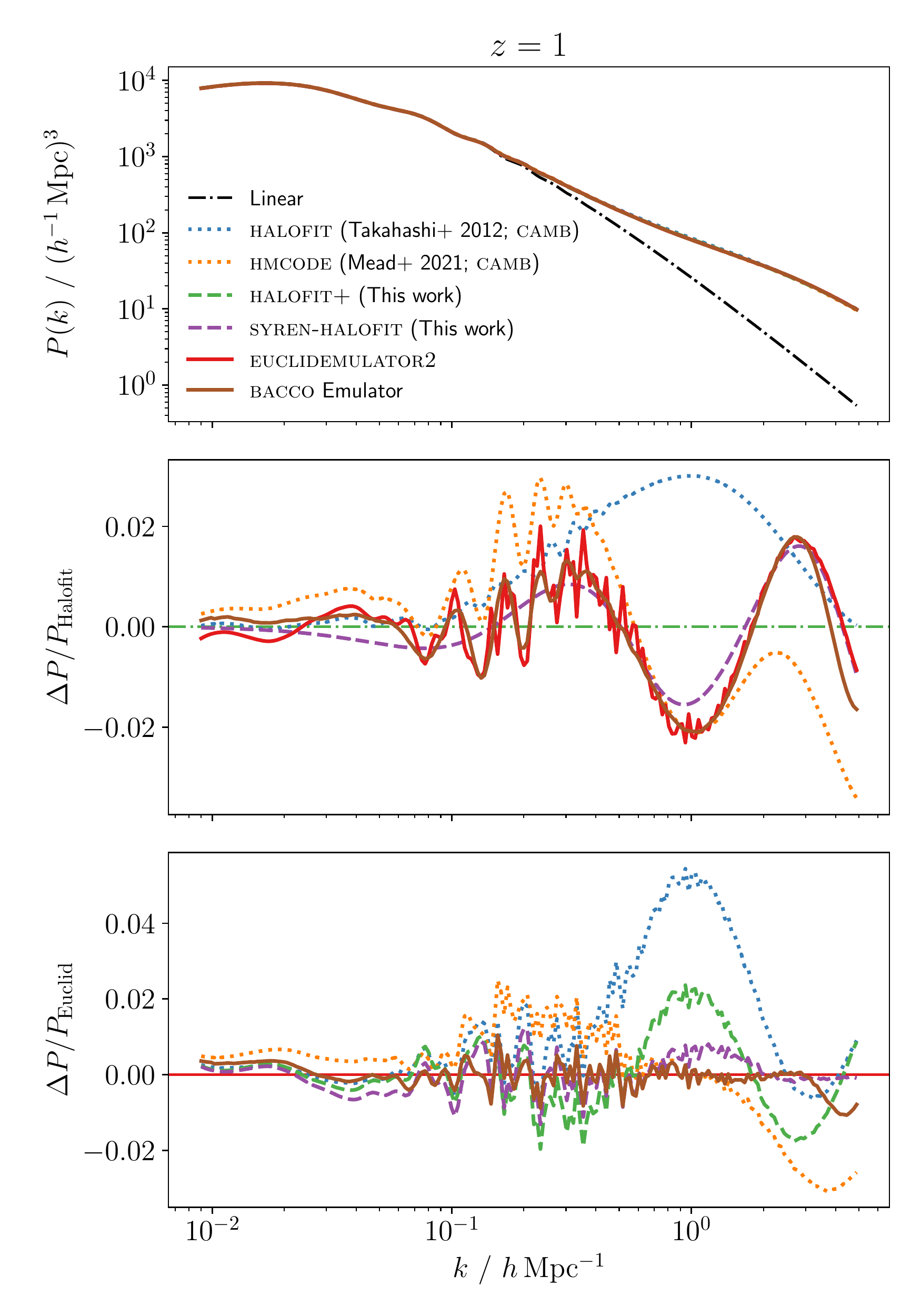}
    \caption{Matter power spectrum predictions at redshift 1.0 for the Planck 2018 cosmology \citep{Planck_VI_2018}. We compare the linear theory prediction to that produced by 
    \halofit{} with the parameters of \citet{Takahashi_2012};
    \hmcode{} as described by \citet{Mead_2021};
    the \halofit{} implementation presented in this work both with (\srhalofit) and without (\halofitplus) the correction given by \cref{eq:Halofit correction};
    \euclidemu;
    and the \bacco{} emulator.
    The top panel shows the predicted power spectrum and the lower two panels give the residuals with respect to 
    the \halofit{} version of this work without the correction of \cref{eq:Halofit correction} and \euclidemu.}
    \label{fig:planck_fit}
\end{figure}

When compared to \cref{fig:new_halofit_errors}, we see that the correction predominantly adjusts the prediction at large $k$, with the two-halo regime being almost unaffected. 
This can be observed in \cref{fig:planck_fit}, where we plot the predictions from \euclidemu, the \bacco{} \citep{Angulo_2021} emulator, \hmcode, and the three versions of \halofit{} considered in this work 
(the parameters of \citet{Takahashi_2012}; 
\halofitplus;
and 
\srhalofit)
for the Planck 2018 cosmology \citep{Planck_VI_2018} at a redshift of 1.0.
From the central panel, we see that the uncorrected \halofit{} implementation leaves a slowly-varying residual of a few percent with respect to \euclidemu{}, which is due to an incorrect prediction from \halofit{} on small scales. The correction given in \cref{eq:Halofit correction} fits this residual, but tends to zero for small $k$, as discussed above.
We find that the maximum fractional difference between 
\srhalofit{}
and \euclidemu{} in \cref{fig:planck_fit} is 1.3\% which, for context, is only 0.3\% greater than the maximum absolute difference between \euclidemu{} and the \bacco{} emulator.

We note that we have chosen not to further re-optimise the coefficients in  \cref{eq:new_an,eq:new_bn,eq:new_cn,eq:new_gamman,eq:new_nu,eq:new_f1,eq:new_f2,eq:new_f3,eq:new_alphan,eq:new_betan} in conjunction with the parameters of \cref{eq:Halofit correction}. Small gains in accuracy are potentially possible under this procedure, but we choose not to perform this optimisation to make the correction factor easier to use (one does not need to use different \halofit{} coefficients if one chooses to switch on the correction factor). 

\section{Emulator peformance}
\label{sec:Performance}

\subsection{Accuracy}
\label{sec:Accuracy test}

\begin{figure*}
    \centering
    \includegraphics[width=\textwidth]{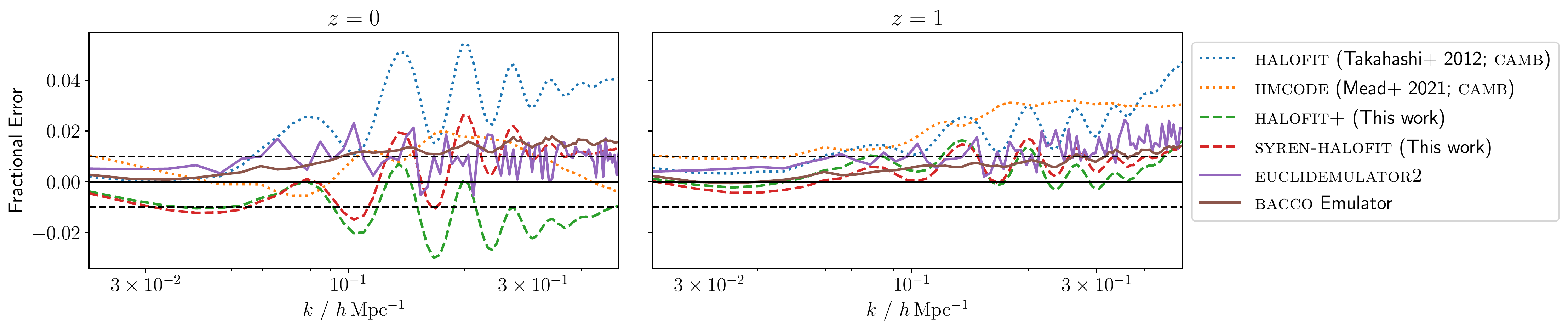}
    \caption{Fractional error on the nonlinear power spectrum as a function of wavenumber, $k$, for various emulators at redshifts 0 and 1. We take the truth to be the average over all \quijote{} simulations considered. For reference, the dashed horizontal lines indicate a $\pm$1\% error band. 
    }
    \label{fig:quijote_comparison}
\end{figure*}

When optimising the parameters of \halofit{} in \cref{sec:Optimise halofit} and obtaining the symbolic correction in \cref{sec:Correct halofit}, we always compared to \euclidemu, since this provided a computationally convenient method for generating large numbers of power spectra at arbitrary redshifts. 
However, it is important to verify that our expressions are accurate when compared to the true outputs of $N$-body simulations, so in this section we do this for both our expressions and other approaches from the literature.

To make this comparison, we compared our predictions to those from the \quijote{} suite of simulations \citep{Quijote_sims}.
These simulations were run within a cubic box of side length $L = 1 \, h^{-1} {\rm Gpc}$ and $N^3=512^3$ particles using the \textsc{gadget-iii} code \citep{Springel_2005}. 
We analysed the results at the fiducial cosmology which has the Planck 2018 cosmological parameters \citep{Planck_VI_2018}: $\Omega_{\rm m} = 0.3175$, $\Omega_{\rm b} = 0.049$, $h=0.6711$, $n_{\rm s} = 0.9624$, $\sigma_8=0.834$.
We chose this suite since there exist 15,000 simulations at this resolution and cosmology, each with different initial conditions. Thus we can average over the large number of simulations so that we suppress cosmic variance and extract the cosmic mean.
The matter power spectrum was extracted at 77 values of $k$ in the range $0.02-0.5 h^{-1}{\rm Mpc^{-1}}$; this $k$ range was chosen since, for larger $k$, the mean from the 15,000 simulations at this resolution deviates by more than few percent of that obtained from the mean of the 100 simulations ran at higher resolutions ($N^3 = 1024^3$), suggesting a lack of convergence on those scales.

In \cref{fig:quijote_comparison} we plot the fractional residuals between the various emulators and the \quijote{} power spectrum at redshifts 0 and 1. As before, we find that the parameters of \citet{Takahashi_2012} results in deviations much larger than a percent for a wide range of $k$. Although \hmcode{} performs well at $z=0$, we find that it performs worse than \halofit{} at intermediate values of $k$ at a redshift of 1. When using our optimised parameters, we find that the fractional error on the 
\halofitplus{}
prediction is much smaller, although the correction of \cref{eq:Halofit correction} is needed to have a percent-level error for larger values of $k$.
We find that the numerical emulators are of comparable accuracy to \srhalofit.

All the \halofit{} predictions (including \halofitplus{} and \srhalofit)
have oscillatory residuals with respect to the truth around the BAO scale.
This suggests that the \halofit{} model cannot sufficiently capture the changes to the BAO features which occur when moving from the linear to the nonlinear power spectrum. Thus, future work should be dedicated to improving the \halofit{} formalism to better capture these oscillations.
We note that the residuals of the \euclidemu{} contain much higher frequency oscillations than \halofit{}, suggesting that these are due to numerical noise. These are not present in the analytic expressions, since it is difficult to produce such high frequency oscillations at moderate model lengths with the basis functions provided.

\subsection{Speed}
\label{sec:Speed test}

\begin{figure}
    \centering
    \includegraphics[width=\columnwidth]{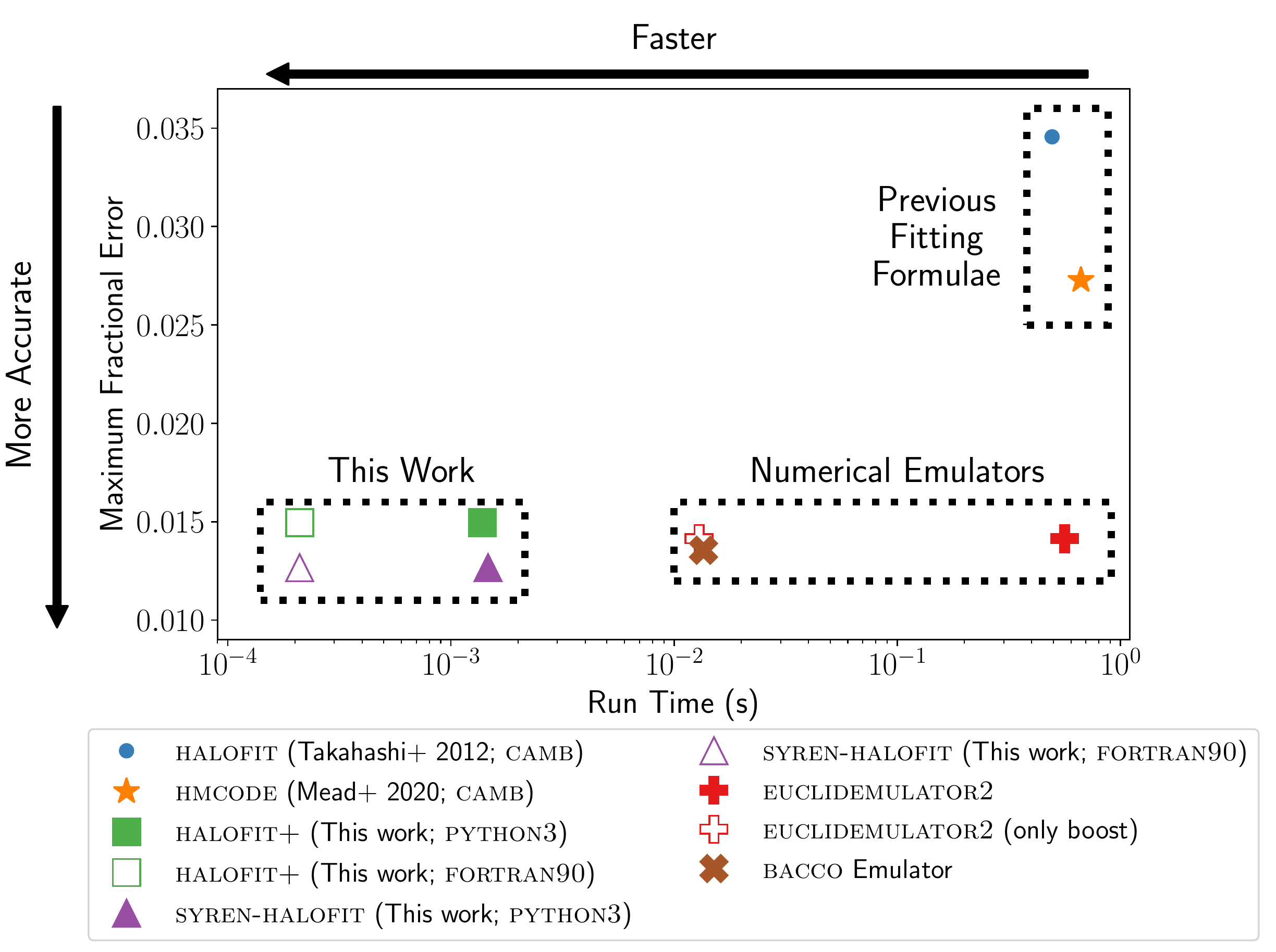}
    \caption{
    Maximum (between $z=0$ and $1$) root mean squared fractional error with respect to the \quijote{} power spectra against run time for various nonlinear power spectrum emulators.
    The run time of \euclidemu{} is dominated by the Boltzmann code \classcode; the emulation run time is comparable to that of the \bacco{} emulator, as indicated by the unfilled cross.
    Our results are orders of magnitude faster than the other methods,
    with our best method -- \srhalofit{} -- achieving even greater accuracy than the numerical emulators.
    }
    \label{fig:timing_test}
\end{figure}

As argued in earlier sections, by avoiding the root-finding procedure and many integrals which need to be performed for the standard implementation of \halofit, we expect that our method should be significantly faster, as well as more accurate.
We compare our approximations (with and without the correction of \cref{sec:Correct halofit}) to the \citet{Takahashi_2012} version of \halofit{} and \hmcode{}, as described in \citet{Mead_2021}, which we compute using \camb{}; \euclidemu{} \citep{Knabenhans_2021}; and the \bacco{} emulator \citep{Angulo_2021}.

To assess the run time, we evaluated the nonlinear power spectrum at redshift 1 for the Planck 2018 cosmology \citep{Planck_VI_2018} $10^3$ times on an 
Intel Xeon E5-4650 CPU, 
and report the mean execution time per evaluation.
We used 200 logarithmically spaced values of $k$ in the range $k=9\times10^{-3}-4.9 \, h{\rm Mpc^{-1}}$.
We note that \camb{} and \euclidemu{} require $A_{\rm s}$, whereas the other methods require $\sigma_8$.
If one had one variable and needed the other, traditional analyses would require one to run a Boltzmann solver to obtain the linear power spectrum to convert between the two. Although \citet{Bartlett_2023_Pofk} produced an analytic expression for this conversion which can bypass this requirement, since it is application-dependent whether one starts with $A_{\rm s}$ or $\sigma_8$, 
we assume that we have available whichever of $A_{\rm s}$ or $\sigma_8$ is required by the emulator in question.
For the implementation of the expressions obtained in this paper, we used the linear matter power spectrum approximation of \citet{Bartlett_2023_Pofk} and considered both a \python{} and a \fortran{} implementation, highlighting the simplicity of exporting symbolic expressions to a different programming language.
In our \python{} implementation we used the \colossus{} \citep{Diemer_2018} package to evaluate the \citet{Eisenstein_1998} transfer function, whereas this was rewritten for the \fortran{} implementation.

The results of this timing test are shown in \cref{fig:timing_test}, where we plot the accuracy -- 
computed as the maximum (between $z=0$ and 1) of the root mean squared fractional errors between the emulator and \quijote{} for the $k$ range used in \cref{sec:Accuracy test}
-- against the run time.
We see that both the corrected (\srhalofit) and uncorrected (\halofitplus) versions of \halofit{} presented in this work require less than $1.5 {\rm \, ms}$ to evaluate in \python{} and just 
$200 {\rm \, \mu s}$ in \fortran, which is significantly shorter than the other methods.
The compilation of the \fortran{} version, itself rapid, only needs to be done once for all further applications of the emulator.
We find that our methods, although symbolic like the implementations of \halofit{} and \hmcode{} in \camb, are a factor of 2350 and 3170 times faster, respectively, when using our \fortran{} code.
Given the short expressions, our method is also faster to evaluate than the the \bacco{} emulator by a factor of 64.

A somewhat surprising feature of \cref{fig:timing_test} is the long runtime of \euclidemu{}; our methods are 2680 times faster than this approach. This becomes less surprising when one realises that \euclidemu{} only predicts the ratio between the nonlinear and linear matter power spectrum (boost) and it relies on using a Boltzmann solver -- in this case \classcode{} \citep{Blas_2011} -- to evaluate the linear power spectrum.
We find that the emulation of the boost takes an average of $13 {\rm \, ms}$ (as indicated by the hollow cross in \cref{fig:timing_test}) which, although much smaller than the total runtime, is still 62 times slower than the symbolic methods we developed.

\section{Conclusions}
\label{sec:Conclusions}

In this paper we have used symbolic regression to eliminate two significant limitations of previous symbolic approximations of the nonlinear matter power spectrum, namely that they were not competitive in terms of speed or accuracy compared to numerical emulators. Our approach is to use symbolic regression with genetic programming as implemented by \operon{} to improve the 
well-established \halofit{} approach
in three ways:

\begin{enumerate}
    \item By providing symbolic expressions for all variables appearing in \halofit, we remove the need to perform integrals, run root-finding algorithms or use a Boltzmann solver. This results in an increase of a factor of 2350 in speed, and is 2680 and 64 times faster than \euclidemu{} and the \bacco{} emulator, respectively. These expressions are given in \cref{eq:ksigma_fit,eq:neff_fit,eq:C_fit}.
    \item We provide updated \halofit{} parameters to fit a wide range of cosmologies between redshifts 0 and 3 designed to provide the best possible predictions for cosmological parameters of interest for present-day studies. We find that this improves the accuracy by approximately a factor of two for $k=9\times10^{-3}-9 \, h{\rm Mpc^{-1}}$. The updated coefficients are given in \cref{eq:new_an,eq:new_bn,eq:new_cn,eq:new_gamman,eq:new_nu,eq:new_f1,eq:new_f2,eq:new_f3,eq:new_alphan,eq:new_betan}.
    \item We obtain a simple analytic expression to multiply the newly optimised \halofit{} prediction by, which reduces the mean squared error to 1\%, providing comparable accuracy to state of the art numerical emulators. This expression is given in \cref{eq:A definition,eq:Halofit correction}.
\end{enumerate}

In this work we have restricted our attention to the standard $\Lambda$CDM model, with zero neutrino mass and a constant equation of state for dark energy. Given that current and upgoing cosmological surveys such as \textit{Euclid} \citep{Euclid_2011}, LSST \citep{LSST_2009}, DESI \citep{DESI_2016} and WFIRST \citep{WFIRST_2019} will be able to probe the nature of dark energy and the neutrino mass, in future work we will extend our formalism for both the linear and nonlinear matter power spectrum to include non-trivial equations of states and a non-zero neutrino mass.
We also note that in this work we have not attempted to alter the underlying form of \halofit{}, except for multiplying the final prediction by a correction function
to obtain \srhalofit.
When inspecting these expressions
(\cref{eq:halofit delta q,eq:halofit delta H,eq:halofit delta Hprime})
and in particular the equations for the parameters which enter them
(\cref{eq:new_an,eq:new_bn,eq:new_cn,eq:new_gamman,eq:new_nu,eq:new_f1,eq:new_f2,eq:new_f3,eq:new_alphan,eq:new_betan}),
one may suppose that more accurate fits could be obtained by using symbolic regression to find a fundamentally more accurate parameterisation. In particular, by automating the search over candidate models, one could allow these to vary in more complicated manners and allow a wider range of cosmological parameters to appear in these expressions. We leave such an investigation to future work.

Accelerating the calculation of important quantities in cosmology does not necessarily require the use of numerical emulators such as neural networks, despite their increasing popularity. We have demonstrated here and in \citet{Bartlett_2023_Pofk} that the more traditional ``fitting function'' approach -- augmented by symbolic regression -- can achieve similar (percent level) accuracy but with shorter evaluation times than these emulators.
The interpretability, clear extrapolation behaviour, portability, and longevity of these analytic approximations bode well for their future in cosmological emulation.

\section*{Acknowledgements}

We thank 
David Alonso,
Anirban Bairagi,
Bogdan Burlacu
Lukas Kammerer
and
Gabriel Kronberger
for useful inputs, comments and suggestions.
DJB is supported by the Simons Collaboration on ``Learning the Universe.''
BDW acknowledges support from the Simons Foundation and from the DIM-ORIGINS "Infinity Next" project.
MZ is supported by STFC.
PGF acknowledges support from STFC and the Beecroft Trust.
HD is supported by a Royal Society University Research Fellowship (grant no. 211046).
We made extensive use of computational resources at the University of Oxford Department of Physics, funded by the John Fell Oxford University Press Research Fund, and at the Institut d'Astrophysique de Paris.

For the purposes of open access, the authors have applied a Creative Commons Attribution (CC BY) licence to any Author Accepted Manuscript version arising.

\section*{Data availability}

The data underlying this article will be shared on reasonable request to the corresponding author.
We provide \python{} and \fortran{} implementations of \srhalofit{} at \url{https://github.com/DeaglanBartlett/symbolic_pofk}. 

\bibliographystyle{aa} 
\bibliography{references} 

\end{document}